\documentclass[twocolumn, pra, showpacs]{revtex4}
\usepackage{graphicx}
\usepackage{amsmath}

\begin{document}

\title{Continuous vortex pumping into a spinor condensate with magnetic fields}
\author{Z. F. Xu,$^{1}$ P. Zhang,$^{2,\dag}$ C. Raman,$^{2}$ and L. You$%
^{1,2}$} 
\affiliation{$^1$Center for Advanced Study, Tsinghua
University, Beijing 100084, People's Republic of China} \affiliation{$^2$School of
Physics, Georgia Institute of Technology, Atlanta, Georgia 30332, USA}

\begin{abstract}
We study the mechanisms and the limits of pumping vorticity into a
spinor condensate through manipulations of magnetic (B-) fields. We
discover a fundamental connection between the geometrical properties
of the magnetic fields and the quantized circulation of magnetically
trapped atoms, a result which generalizes several recent
experimental and theoretical studies. The optimal procedures are
devised that are capable of continuously increasing or decreasing a
condensate's vorticity by repeating certain two step B-field
manipulation protocols. We carry out detailed numerical simulations
that support the claim that our protocols are highly efficient,
stable, and robust against small imperfections of all types. Our
protocols can be implemented experimentally within current
technologies.
\end{abstract}


\maketitle

\section{Introduction}
A quantized vortex represents a hallmark of superfluidity
\cite{leggett}. The vorticity in a superfluid
clearly reveals the topological nature associated with the phase of a
condensate's wave function and has been studied extensively both
theoretically and experimentally \cite{leggett,fetter} since the
first success of atomic Bose-Einstein condensation. In the
strongly rotating limit when the number of vortices significantly
exceeds the number of atoms, researchers have focused
on the possibilities of observing interesting
strongly correlated states \cite{papen,kav1,kav2,
zol,hofr,wilkin,mac,baym,baranov,jila0,ens0,mit0,ryu,morris}.

Vortex states have been created experimentally relying on a variety of
approaches: a spatially selective rotating Rabi coupling to
an auxiliary internal state \cite{jila1,jilat}, stirring inside \cite{mit}
or at the edge \cite{ens} of a condensate with a focused laser beam,
rotating a deformed trap \cite{eng}, or
simply merging several condensate pieces \cite{col}
into a single trap. Other approaches include direct phase engineering
through the imprinting of a nontrivial topological phase
to the wave function of a condensate \cite{ertmer}
using an azimuthal optical phase plate \cite{nist} or through
manipulating external magnetic (B-) fields \cite{Machida,p1,p3,p4}.
Alternately, a condensate can gain angular momentum
through its coupling to a Laguerre-Gaussian light beam \cite{wp,rm1,rm2,ulm}
or other forms of gauge potentials \cite{gpt}.

The idea of phase imprinting from the nontrivial geometrical
properties of external B-fields is especially
illuminating as it involves no forced rotation or essentially
no spatial motion \cite{Machida}.
It has since been demonstrated in beautiful recent experiments
\cite{emit1,emit2,japan}.
The original protocol \cite{Machida} involves an axially symmetric
spinor condensate in the familiar B-field of
an Ioffe-Pritchard trap (IPT) \cite{Pritchard}
consisting of a two dimensional (2D) quadrupole B-field
augmented by a bias B-field along the symmetric $z$-axis.
Upon the flipping of the bias field, the adiabatically following
weak field seeking state then develops a nonzero vorticity
in its spatial wavefunction \cite{Machida}.
Our aim is to develop this protocol into a pump mechanism
capable of continuously changing
the vorticity of a condensate. The naive approach of
repeated flips of the $z$-bias B-field does not work because the
second flip simply undoes the vorticity gained in
the first flip, returning the condensate to the initial state.

Recently, M\"ott\"onen, {\it et al.} \cite{pump}, put forward
an interesting idea that sought to break the time reversal symmetry between
the first and the second flips of the bias field. Their protocol
starts with a 2D hexapole B-field instead of a quadrupole
field for the first flip. The hexapole field is then turned off
and replaced by a quadrupole one for the
second flip. It is known, as in the original bias flip
protocol \cite{Machida}, that the first flip can generate a proportionally higher
vorticity if the quadrupole is replaced by a higher order
2D multipole such as a hexapole \cite{aaron}. Repeating the two
step protocols with hexapole and quadrupole B-fields in turns,
M\"ott\"onen, {\it et al.}, show that the
vorticity increases by $4\hbar$ per atom in the hexapole step and
decreases by $2\hbar$ per atom during the quadrupole step
\cite{pump}. Thus each cycle composed of a bias flip with a hexapole
field followed by a second bias flip with a quadrupole field increases the
net vorticity by $4\hbar-2\hbar=2\hbar$. Higher vorticities are
generated with repeated cycles.

A related earlier discussion \cite{aaron} suggested an even simpler
protocol of turning off the quadrupole field after the first flip,
and then returning the axial bias field to its original direction.
The return of the axial bias to the original direction is hoped to
bring the system to the initial configuration except for the
vorticity gained from the first flip. It was previously noted
that by turning on the quadrupole field
and repeating the above protocol, a continuous vortex pump is realized
\cite{aaron}. Unfortunately, no details were given on how the axial
bias field is returned to the original direction \cite{aaron}.
Clearly it cannot be flipped back in the absence of the quadrupole
field as the $z$-bias field remains aligned along the same axial
direction. Simply let it oscillate back with the axial B-field
magnitude increasing from large negative to small negative, to zero,
to small positive, and then finally to large positive values does
not constitute a flip. Both the projections of mechanical angular
momentum $L_z$ and the spin $F_z(M_F)$ are independently conserved
quantities during the above process.

\section{Our continuous vortex pump protocol}

Our idea is physically intuitive and leads to a direct
implementation for a continuous vortex pump with the introduction of
an auxiliary transverse bias B-field. As in a Stern-Gerlach
experiment, the Zeeman population distribution inside an
experimental apparatus can be measured along any quantization
direction of a strong reference B-field, provided the strong field
is turned on adiabatically. When the transverse bias B-field is
pulsed on, it provides a reference direction for the axial bias
field to flip back in the absence of the quadrupole field. We find
it imperative to present this result because of the high interests
stimulated by the recent work of M\"ott\"onen, {\it et al.}
\cite{pump}. Unlike the Ref. \cite{pump}, our proposal requires only
one set of current coils capable of generating one type of multipole
B-field: be it a 2D quadrupole or a 2D hexapole. The time reversal
symmetry is broken in our protocol with the extra transverse bias
B-field whose presence allows for the $z$-bias field to rotate instead
of simply oscillating back.

As an illustration we consider a spin-1 condensate in an IPT, whose
B-field is approximately $\vec
{B}(x,y,z)=B'(x\hat{x}-y\hat{y})+B_z\hat{z}$ near the origin.
Isoshima {\it et al.} \cite{Machida} first discussed
pumping vorticity into a condensate with external B-fields.
For a sufficiently large $z$-bias B-field, a condensate of $F=1$ atoms
adiabatically stays in the B-quantized $|M_F=-1,\vec{r}\rangle_B$
state. After an adiabatic flip of the bias field, a vortical phase
structure or a vortex state is imprinted into the condensate
\cite{p1,p3,p4,emit1,emit2,japan}. The physics involved can be
elucidated in terms of the symmetries and conservation laws of the
model system \cite{peng} or from the gauge potential \cite{Ho} due
to the changing B-field. More specifically, it is the conservation
of $D_{z}$ ($=L_z-F_z$) that enables a straightforward understanding
of the result of Isoshima, {\it et al.} \cite{Machida}, in the
axially symmetric IPT \cite{hinds,pot}. Each single flip of the bias
field in the presence of a quadrupole field then imparts a $2F\hbar$
(for $F=1$ here) phase winding \cite{Machida,emit1,emit2,japan}.
Repeated operations then constitute a continuous vortex pump
\cite{pump}. As introduced above, our pump mechanism
generalizes the
original idea of Isoshima, ${\it et\ al.}$ \cite{Machida}. Following
the first flip, which imprints a $2\hbar$ vortex to the adiabatic
state, we turn on a transverse bias field $B_x^r\hat{x}$ and then continue with a
second bias flip returning the system to the initial setup. A vortex
pump then simply consists of repeated applications of the above
manipulations to the three B-fields: the 2D quadrupole, the axial
bias, and the transverse bias fields.

The potential for high fidelity operation of our pump protocol is
confirmed with numerical simulations for a spin-1 atomic condensate
in external magnetic (B-) fields. An additional optical trap
$V_o=M\omega_{\perp}^2(x^2+y^2+\lambda^2z^2)/2$ provides a permanent
confinement during the B-field manipulation. To avoid the energetic
instability of the potential disintegration of a high vortex state
into single vortices \cite{t1,t2,t3}, we introduce an optical
pinning plug $V_{p}(\rho,z,\phi)=U\exp(-\rho^2/2\rho_0^2)$ that
expels atoms away from the low B-field region to assure
adiabaticity. The force of gravity is assumed to be opposite to the
axial $z$-axis, which can be omitted in the approximate 2D treatment
when the harmonic trap is taken to be pancake shaped with
$\lambda\gg1$. The $z$-dependence is frozen in the ground state of
the axial harmonic oscillator $\phi(z)$. Thus the vorticity is coded
in the phase structures of the two dimensional condensate wave
function $\psi({\vec\rho},t)$.

The condensate wave function thus becomes
$\Psi(\vec{r},t)=\psi({\vec\rho},t)\phi(z)$, where $\vec\rho=(x,y)$,
$\phi(z)=(\lambda/\pi a^2_{\perp})^{1/4}e^{-\lambda
z^2/2a_{\perp}^2}$, and $a_{\perp}=\sqrt{\hbar/M\omega_{\perp}}$ is
the length scale for the transverse harmonic trap. After integrating
out the $z$ coordinate, we obtain the effective 2D Gross-Pitaevskii
equation
\begin{eqnarray}
    i\hbar\frac{\partial\psi_{\pm1}}{\partial t}&=&
    \left[\frac{}{}H_0+H_{\pm1\pm1}^{ZM}+c_2^{(2D)}(n_{\pm1}+n_0-n_{\mp1})\right]\psi_{\pm1}\nonumber\\
    &&+c_2^{(2D)}\psi_{\mp1}^*\psi_0^2+H^{ZM}_{\pm10}\psi_0+H^{ZM}_{\pm1\mp1}\psi_{\mp1},\nonumber\\
    i\hbar\frac{\partial\psi_0}{\partial
    t}&=&\left[\frac{}{}H_0+H_{00}^{ZM}+c_2^{(2D)}(n_1+n_{-1})\right]\psi_0\nonumber\\
    &&+2c_2^{(2D)}\psi_0^*\psi_1\psi_{-1}
    +H^{ZM}_{01}\psi_1+H^{ZM}_{0-1}\psi_{-1},\hskip 18pt
    \label{gpe}
\end{eqnarray}
where
$H_0=-\frac{\hbar^2}{2M}\nabla^2_{\perp}+V_o^{2D}+V_p^{2D}+c_0^{2D}n$,
$V_o^{2D}=\frac{1}{2}M\omega_{\perp}^2(x^2+y^2)$ and
$V_p^{2D}=U\exp(-\rho^2/2\rho_0^2)$ are the 2D optical trap and
optical plug, respectively. $n=\sum_i|\psi_i|^2$ is the 2D atom
number density. The effective 2D spin-independent and spin-dependent
interaction strengths are now characterized by
$c_0^{(2D)}=2(2\pi\lambda)^{1/2}\hbar^2(a_0+2a_2)/3Ma_{\perp}$, and
$c_2^{(2D)}=2(2\pi\lambda)^{1/2}\hbar^2(a_2-a_0)/3Ma_{\perp}$.

In the local B-quantized representation, the Zeeman energy is
diagonal, given by the Breit-Rabi formula
\begin{eqnarray}
    E_{M_F}=-\frac{\Delta E_0}{8}-M_Fg_I\mu_IB-\frac{\Delta
    E_0}{2}\sqrt{1+M_F\xi+\xi^2}\,.\hskip 12pt
\label{breit-rabi}
\end{eqnarray}
$\Delta E_0$ is the hyperfine splitting. $g_I$ is the Lande factor
for the nuclear spin $\vec{I}$, $\mu_I$ is the nuclear magneton, and
$\xi$ is defined by $\xi=(g_I\mu_I+g_J\mu_B)B/\Delta E_0$. Here
$g_J$ is the Lande factor for the valence electron with total
angular momentum $\vec{J}$, and $\mu_B$ is the Bohr magneton. Our
equation (\ref{gpe}) above requires Zeeman energy $H^{ZM}$ of an
atom in the laboratory based $z$-axis quantization representation
that is nondiagonal and is given by $H^{ZM}={\cal U}^{\dag}E{\cal
U}$ with ${\cal U}$ the unitary transformation connecting the $z$-
and B-quantized representations. For the B-field of a IPT, the
transformation matrix becomes the product of two rotations in spin
space, i.e., ${\cal U}=e^{iF_y\theta}e^{iF_z\phi}$, or in matrix
form,
\begin{eqnarray}
    {\cal U}=\left(
    \begin{array}{ccc}
    \frac{1}{2}(1+\cos\theta)e^{i\phi} & \frac{1}{\sqrt{2}}\sin\theta & \frac{1}{2}(1-\cos\theta)e^{-i\phi} \\
    -\frac{1}{\sqrt{2}}\sin\theta e^{i\phi} & \cos\theta & \frac{1}{\sqrt{2}}\sin\theta e^{-i\phi} \\
    \frac{1}{2}(1-\cos\theta)e^{i\phi} & -\frac{1}{\sqrt{2}}\sin\theta & \frac{1}{2}(1+\cos\theta)e^{-i\phi} \\
    \end{array}
    \right),\ \
\label{unitarytransformation}
\end{eqnarray}
with the rotation angles $\theta$ and $\phi$ introduced through the
parametrization of the field
 $\vec B=B[\hat{z}\cos\theta
+\sin\theta(\hat{x}\cos\phi +\hat{y}\sin\phi)]$. $H^{ZM}_{M_FM_F'}$
is one of the nine elements of $H^{ZM}$ with spinor components
ordered as $M_F=1,0,-1$.

\begin{figure}[tbp]
\centering
\includegraphics[width=3.4in]{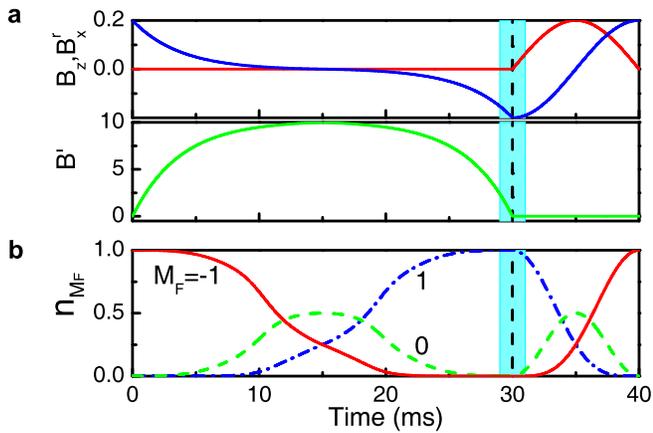}
\caption{(Color online). (a) The time dependence of the B-fields. In
the first 30 ms, $B_z$ decreases to the opposite of the initial
value. The gradient increases from $0$ to the maximum in 15 ms, then
decreases to zero. Next, the B-field is rotated back, while a
uniform transverse bias in the $x$-axis direction is pulsed on for
the latter 10 ms, keeping the $B_z^2+(B_x^r)^2$ a constant. (b) The
time dependent fractional population $n_{M_F}=N_{M_F}/N$ in the
Zeeman state.} \label{fig1}
\end{figure}

Our numerical simulations assume a condensate of $N=10^5$, $^{87}$Rb
atoms in the $F=1$ state and trapped as described above with
$\lambda=100$. We take $\omega_{\perp}=2\pi\times 30$ Hz,
$U/\hbar=2\times10^5$ Hz, and $\rho_0=5$ $\mu m$. The B-field takes
a form $\vec B=B'(x\hat{x}-y\hat{y})+B_z\hat{z}+B^{r}_x\hat{x}$ with
a typical temporal evolution within one period as illustrated in
Fig. \ref{fig1}. During the first $15$ ms, we decrease $B_z$ and
increase $B'$ according to the following easily programmed time
dependence
\begin{eqnarray}
    B_z(t)&=&B_z(0)(e^{-8t/T_1+2}-e^{-2})/(e^{2}-e^{-2}),\\
    B'(t)&=&B'(T_1/2)(e^{2}-e^{-8t/T_1+2})/(e^{2}-e^{-2}),
\label{Bzpt1}
\end{eqnarray}
with $T_1=30$ ms. In the subsequent $15$ ms, $B_z$ is
 decreased continuously as is the gradient $B'$.
Both time dependence are relatively smooth and of an exponential
type similar to the first $15$ ms. The time dependence of the
$z$-bias flip and the transverse $x$-bias pulse on are described by
\begin{eqnarray}
    B_z(t)&=&B_z(0)\cos[\pi(1-(t-T_1)/T_2)],\\
    B^r_x(t)&=&B_z(0)\sin[\pi(1-(t-T_1)/T_2)],
\label{Bzpt2}
\end{eqnarray}
with $T_2=10$ ms. Due to the adiabatic nature of our proposed
protocol, the quality of the final result or the fidelity of the
intended vortex state do not sensitively depend on the details of
the time dependence of the B-fields, provided they are reasonably
smooth functions of time. In the shadow window of Fig. \ref{fig1},
nothing really happens to the condensate spinor component population
distribution as long as the B-fields change smoothly. Because $B_z$
is much larger than the field due to the gradient $B'$ within this
window, the angle between the net B-field and the $z$-axis remains
small even for a discontinuous change of the bias field or the
quadruple gradient, causing the atomic spin state to change
smoothly. Depending on the net radial trap strength from the
combined trap and the optical plug, the azimuthal vortical phase
distribution changes with the radius due to non-adiabatic effects
with respect to the radial motional state.

\begin{figure}[tbp]
\centering
\includegraphics[width=3.4in]{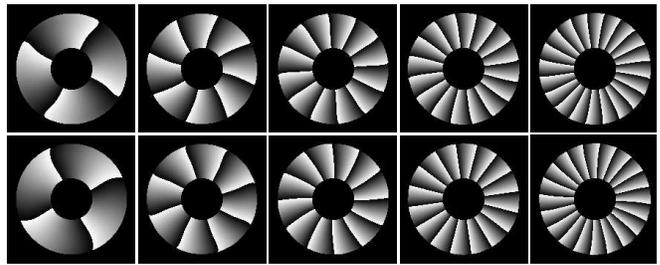}
\caption{(Color online). The temporal development of the condensate
phase structures in a hexapole field
$\vec{B}_h=B'_h[(x^2-y^2)\hat{x}-2xy\hat{y}]$ plus a transverse bias
field $B_x^r\hat{x}$. The upper row shows the $M_F=1$ component at
the end of the first flips, or in the middle of each cycle at $30$,
$70$, $110$, $150$, and $190$ ms, respectively; the lower row is for
the $M_F=-1$ state at $10$ ms later with respect to the first row or
at the end of each cycle at $40$, $80$, $120$, $160$, and $200$ ms.
The temporal dependence of the B-fields and other parameters are as
in Fig. 1 of the supplementary material. The maximum for $B'_h$,
$B_z$, and $B_x^r$ are $1250$ Gauss/cm$^2$, $0.1$, and $0.1$ Gauss,
respectively. White (black) color denotes phase $-\pi$ ($\pi$). }
\label{fig2}
\end{figure}

In all of our simulations, the initial state is obtained from the
imaginary time propagation of the coupled Gross-Pitaevskii equations
with an initial Gaussian wave function in the $M_F=-1$ state setting
$c_2=0$ and with no B-field as discussed in the supplementary
material. The vortex pump protocol is then simulated in real time.
For the case of a 2D quadrupole,
the first flip of the bias $B_z$ from $0.2$
Gauss to $-0.2$ Gauss is carried out in the first $30$ ms,
along with the increasing of
the B-field gradient $B'$ to the maximum value of $10$ Gauss/cm and
followed with a gradual switch off. In the following $10$ ms the
bias field is rotated back around the $y$-axis in the $x$-$z$ plane,
aided by the transverse bias field along the $x$-axis.
Similar steps are involved when the 2D quadrupole is replaced
by a 2D hexapole.
In Fig. \ref{fig2} we show the temporal development of the phase structures
for the $M_F=1$ component after the first $B_z$ flip and the
$M_F=-1$ component after the next flip back to the original
direction for the first five repetitions. The total vorticity
changes are impressive
and confirm conclusively our suggested protocol for a continuous
vortex pump applied with a hexapole. For a quadrupole field, we find
essentially the same quality operation, except that each repetition
adds $2\hbar$ units of vorticity as compared to $4\hbar$ units for a
hexapole (Fig. \ref{fig2}). More generally with a $z$-bias field
along the symmetry axis of a $2l$-th multipole 2D B-field
$\vec{B}\propto\rho^{l-1}[\cos(l\phi)\hat{\rho}-\sin(l\phi)\hat{\phi}]$,
$D_z=L_z-(l-1)\times F_z$ is conserved for the symmetric ground
condensate state \cite{peng}. A corresponding adiabatic flip of the
bias field is thus capable of generating a $2(l-1)F\hbar$ vortex.

Before providing the general considerations that will
lead to our discovery of the optimal vortex pump
protocols based on manipulating external B-fields,
we provide further numerical studies that demonstrate
the efficiency, stability, and robustness of our vortex pump protocol.

\begin{figure}[tbp]
\centering
\includegraphics[width=3.4in]{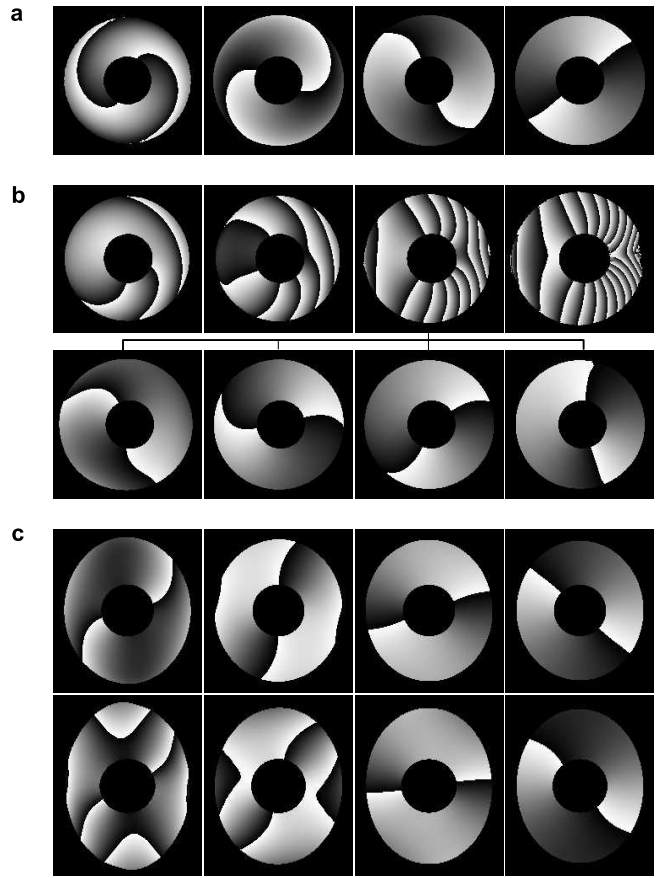}
\caption{(Color online). The condensate phase distribution for the
$M_F=1$ state after flipping the $z$-bias in an IPT. (a) The
ideal case of an IPT with perfect alignment with the optical trap
and the plug. The flip times are 30, 60, 90, and 120 ms from the
left to the right, accompanied by an increasing level of adiabaticity.
(b) The same as in (a) except for a misaligned optical plug
$V_p^{2D}=U\exp(-((x-x_0)^2+y^2)/2\rho_0^2)$, $U/\hbar=2\times 10^5$
Hz, and $\rho_0=5\,\mu m$. From the left to the right in the upper
row the results are compared for $x_0/\rho_0=0.05, 0.1, 0.2$, and
$0.3$, respectively, at a bias flip time of 30 ms. In the lower row
$x_0=0.2 \rho$, where results from different flip times of 60, 90,
120, and 150 ms are compared from the left to the right. (c) The
same as in (a) except for a non-axial-symmetric optical trap
$V_o^{2D}=M\omega_{\perp}^2[
(1+\varepsilon)^2x^2+(1-\varepsilon)^2y^2]/2$ parameterized by
$\varepsilon$. The upper row is for $\varepsilon=0.05$, and
$\varepsilon=0.1$ in the lower row. The $z$-bias flipping times used
are 60, 90, 120, and 210 ms from left to right, respectively.
} \label{fig3}
\end{figure}

First, we discuss the relatively relaxed conditions on adiabaticity
for the manipulation of the external B-field. Typically, the
$z$-bias field is flipped over during a time of 30 ms in our
numerical simulations, although we find integer numbers of vorticity
are still created provided the flipping time is adjusted
considerably (shorter or longer).
Due to the symmetries
and the corresponding conservation laws on the dynamics of our model
system; the conservations of $D_z=L_z-F_z$ in an IPT, $L_z+F_z$ in a
3D quadrupole trap (QT) and $L_z-(l-1)F_z$ in a 2l-th multipole 2D
B-field, high fidelity operations are assured even with
marginal conditions for adiabaticity.
On careful examination, we find that
the aximuthal distributions of the phases are complicated, not the
simple linear dependence one might have imagined.
Instead, the dependence on the
azimuthal angle is different at a different radial coordinate,
indicating a violation of the adiabatic condition with respect to
the radial motional state. However, as long as the internal state
can follow adiabatically the flipping $z$-bias B-field, our
intuitive vortex pump protocol remains effective. As we demonstrate
in Fig. \ref{fig3}(a), for successive flipping times of 30 ms to 60,
90, and 120 ms, the phase distribution approaches the same form with
the increasing level of adiabaticity.

Next, we study the effect of misalignment of the axial symmetric
optical trap with respect to the geometric center of the external
B-field. Any misalignment will reduce the axial symmetry,
and thus break the angular momentum conservation law \cite{peng} and
adversely impact the quality of our vortex pump protocol. Without
loss of generality, we assume the optical plug's center is $(x_0, 0,
0)$ while other symmetries are assumed to remain. After extensive
numerical simulations as before, we find that a misalignment of the
percent level of the transverse size for the condensate is
already detrimental to the pump protocol as shown in the upper row
of Fig. \ref{fig3}(b) when the bias flip time is 30 ms.
More generally, we find that the larger is $x_0$, the stronger
is the deviation from the intended vortical phase distribution pattern.
Somewhat surprisingly, however, we find a simple solution to this
ailment. By increasing the $z$-bias flip time, the complicated phase
structures disappear gradually as shown in the lower row of Fig.
\ref{fig3}(b), where the flip time is taken to be 60, 90, 120
and 150 ms from left to right. These results show that unless there is
a total failure in aligning the optical trap with respect to
the B-field trap, the nominally small misalignment can be
compensated for by the proposed vortex pump protocol with a
correspondingly slower flipping time.

\begin{figure}[tbp]
\centering
\includegraphics[width=3.4in]{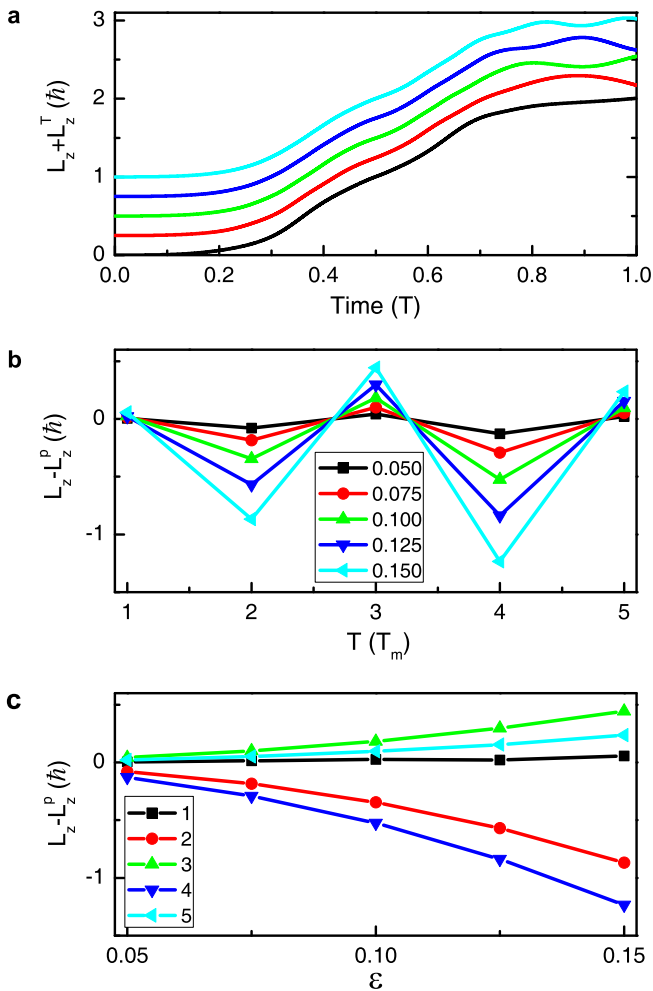}
\caption{(Color online). The motional angular momentum gained
with our protocol in an asymmetric optical trap and
an IPT. (a) The temporal development of $L_z$ during the
flip of the $z$-bias field at an asymmetry
parameter $\varepsilon=0.05$. The flip time $T$
is 30, 60, 90, 120, and 150 ms,
respectively, and is shown by off-set lines from the lower to the higher.
The off-set shift for each curve is given by
$L_z^T=0.25\hbar\times$[($T/T_{m}$)-1], and $T_m=30$ ms. (b)
The difference between angular momentum gained from
our protocol in an asymmetric optical trap with
$L_z^p=2\hbar$ predicted for the ideal case.
The curves denote different asymmetry parameter
$\varepsilon$ from 0.05 to 0.15 as shown by the insert labels.
(c) The same as in (b), but now for the $\varepsilon$
dependence at the same flip time T. The five curves are labelled
by their respective $T/T_m$ values from 1 to 5.}
\label{fig4}
\end{figure}

Finally, we study the potential degradation due to an asymmetric optical
trap. As with the misalignment considered above, an asymmetric
optical trap will void the conservation of $D_z$. Although the
intended adiabatic manipulation of the B-field can still transform
all atoms from the $M_F=-1$ to $M_F=1$ spin state and cause the
$F_z$ to change from $-\hbar$ to $\hbar$ for a spin-1 condensate,
the axial motional angular momentum per atom gained will not be
equal to $2\hbar$, except when the asymmetry is small. Instead,
$L_z$ generally is found to oscillate near the end of the B-field
flipping, as we show in Fig. \ref{fig4}. For relatively small trap
anisotropy, for instance at $\varepsilon=0.05$, the vortex generated
is not far away from the ideal case of perfect symmetry.
This can be readily derived from the axial angular momentum shown in
Fig. \ref{fig4}(a) or by comparing the condensate's phase structure in Fig.
\ref{fig3}(c) with Fig. \ref{fig3}(a). With larger optical
trap asymmetries, for example at the threshold of $\varepsilon=0.1$
for the system and parameters that we consider,
 the average angular momentum per particle ($L_z$)
 gained by the condensate is no longer an integer
multiple of $\hbar$. The condensate phase structure becomes more
complicated at short flip times, and if the optical trap asymmetry
is distorted further a dynamical instability begins to set in within
the same evolution time. Based on our numerical simulations, when
$\varepsilon=0.125$ and the flipping time is set to 30 ms, a
dynamical instability occurs. In addition, we find significant
blurring in the condensate density distribution after flipping the
B-field. The phase difference between nearby spatial
positions is totally smeared out.
As the flipping time is increased further,
dynamical instability or the blurring disappears
while high quality winding pattern reemerges.
Our simulations also show that the dynamical instability
survives at larger rotations for smaller trap asymmetries.

\section{Geometrical properties of static magnetic fields}

According to Maxwell's equation, the spatial distribution of a
static B-field can be expressed as the
gradient of a scalar potential $\Phi (\vec{r})$, {\it i.e.},
\begin{eqnarray}
\vec{B}(\vec{r})=[B_{x}(\vec{r}),B_{y}(\vec{r}),B_{z}(\vec{r})]
=\nabla \Phi (\vec{r}).
\end{eqnarray}%
Since the divergence of a static B-field is always zero, or
$\nabla\cdot \vec B=0$, the scalar potential satisfies the laplace
equation $\nabla^2 \Phi (\vec{r})=0$. Thus $\Phi (\vec{r})$ can be expanded in
terms of the spherical harmonic functions, to
\begin{eqnarray}
\Phi (\vec{r})=\sum_{lm}c_{lm}r^{l}P_{l}^{m}(\cos \theta )e^{im\phi},
\label{sp}
\end{eqnarray}%
in the spherical coordinate $(r,\theta,\phi)$.

The vortex pump protocol we proposed relies on the repeated flips of
the $z$-bias field $B_{z}$ adiabatically from the $+z$ to the $-z$
direction and vice versa. The vorticity gained by a condensate can
be understood in terms of the conservations of $D_z$ or $L_z+F_z$,
which are ultimately
determined by the angle $\vartheta$ between $\vec{B}%
(\vec{r})$ and the $x$-axis (or the $y$-axis). A
vortex state has a two-dimensional phase structure.
For a stable vortex state with its angular momentum pointed along
the $z$-axis, the velocity
$\vec{v}=\frac{\hbar}{M}\nabla\phi_p$ must flow
along the same azimuthal direction, and the phase $\phi_p$ in any plane
with $z=z_0$ must be similar to that in the $z=0$ plane,
{\it i.e.}, $\phi_p(\rho,z,\phi)=\phi_p(\rho,0,\phi)+f(z)$.
To effectively obtain a stable vortex state,
the angle $\vartheta$ should (i)
depend only on $\phi$ but remain independent of $z$,
{\it i.e.}, $\vartheta(\rho,z,\phi)=\vartheta(\rho,0,\phi)$;
or (ii) depend on both $\phi$ and $z$
but take the form
$\vartheta(\rho,z,\phi)=\vartheta(\rho,0,\phi)+g(z)$. These
constrains then lead to the condition
\begin{eqnarray}
c_{lm}=0,
\end{eqnarray}
where $m\neq \pm l$ or $m\neq 0$ on the expansion coefficients. To
obtain a vortex state with a definite winding number, the scalar
potential thus has to take one of the following two possible forms
\begin{eqnarray}
\Phi_{l}^{(I)}(\vec{r})\propto \rho^{l}\cos (l\phi+\eta),
\end{eqnarray}%
or
\begin{eqnarray}
\Phi^{(II)}(\vec{r})=\sum_{l}c_{l}r^{l}P_{l}^{0}(\cos\theta),
\end{eqnarray}%
where $(\rho,z,\phi )$ are the cylindrical coordinates and $\eta$
is a constant angle. In the derivation of the second form,
we have used the fact that $\Phi (\vec{r})$ is real.

In the first case above and as was discussed in the thesis work of
A. E. Leanhardt \cite{aaron}, the $B$-field takes the form
\begin{eqnarray}
\vec{B}_{l}^{(I)}(\vec{r})\propto \rho ^{l-1}\left[ \cos (l-1)\phi \hat{e}%
_{x}-\sin (l-1)\phi \hat{e}_{y}\right],
\end{eqnarray}%
with  $l\geq 2$. The corresponding conserved quantity becomes
$D_z=L_z-(l-1)\times F_z$. When $l=2$, this is the field of an IPT.
In the work of Mikko M$\ddot{\rm o}$tt$\ddot{\rm o}$nen \cite{pump},
both fields $\vec{B}_{3}^{(I)}(\vec{r})$ of a hexapole and
$\vec{B}_{2}^{(I)}(\vec{r})$ of a quadrupole are used. For a
spin-$1$ condensate, the vorticity gained is $L_z=2(l-1)\hbar$ after
adiabatically flipping the $B_{z}(\vec{r})$.

In the second case, the simplest B-field is
\begin{eqnarray}
\vec{B}^{(II)}(\vec{r})=B_{\rho }(\vec{r})\hat{e}_{\rho }+B_{z}(\vec{r})\hat{e%
}_{z},
\end{eqnarray}%
as in a 3D QT. In the $x$-$y$ plane, $\vec{B}^{(II)}$ points along
the direction of $\hat{e}_{\rho}$. For a spin-$1$ condensate,
the vorticity gained is $L_z=-2\hbar$ per atom after the
flipping of $B_{z}(\vec{r})$. For completeness, we have confirmed
this numerically for a 3D quadrupole trap. A slight complication
arises in this case, as the B-field depends on the $z$-coordinate,
and its zero value point moves from $-\infty$ to $+\infty$ during
the flipping of the $z$-bias from $-0.1$ to $0.1$ Gauss. To
demonstrate the numerics, we used the coupled 3D
Gross-Pitaevskii equations. The gain of a $L_z=-2\hbar$ vortex
also can be understood in this case as being due to the conservation of
$F_z+L_z$ \cite{peng}.

\section{The optimal vortex pump protocol}

As discussed above in some detail, starting in the initial spin
state $|(-1)\rangle_{z}$, vorticity can be pumped into the
motional state of a condensate if the $z$-bias field $B_{z}$ is
flipped from the positive to the negative z-axis.
When the static field is of the form
$\vec{B}_{l}^{(I)}(\vec{r})$, the vortex state has
$L_z=2(l-1)\hbar$;
If the form is changed to $\vec{B}^{(II)}(\vec{r})$, a vortex state
of $L_z=-2\hbar$ is obtained. The geometrical constraint on a static B-field
together with its coupling to the atomic hyperfine spin stops
the change of the atom's internal state through the flipping
    of the $z$-bias B-field. In fact the opposite vortex state
    of $-2(l-1)\hbar$ with $l\geq 3$ can never be created through the
    flipping of $B_z$ associated with any B-field.
Clearly, our studies provide an
optimal vorticity pumping protocol based on
a cyclic flipping of the bias $B_z$ field as devised below.

If we have access only to the field $\vec{B}_{l}^{(I)}(\vec{r})$, we
can turn it on and generate a vortex with $L_z=2(l-1)\hbar$ through
the flipping of the bias in the first step. Then, this is followed by
a second step where the field $\vec{B}_{l}^{(I)}(\vec{r})$ is turned
off, and the bias field $B_z$ is flipped back with the help of a
constant transverse bias $B_{x}^r\hat{x}$ as we have suggested. The
second step does not create any vorticity but returns the internal
state to what it was initially. Repeating the two step protocols,
we pump $L_z=2(l-1)\hbar$ units of vorticity into the condensate for
each cycle. The motional angular momentum gained from this protocol
for the first five repetitions is shown in Fig. \ref{fig6}(a).

If we have access to both fields $\vec{B}_{l}^{(I)}(\vec{r})$ and
$\vec{B}^{(II)}(\vec{r})$, the second step of each cycle is modified
by tuning on the field $\vec{B}^{(II)}(\vec{r})$, which gains
an additional $2\hbar$ units of vorticity in each cycle, for a total of
$2l\hbar$ angular momentum gained per cycle. Figure \ref{fig6}(b)
shows the continuously increasing vorticity for this more general
case of the two type 2D multipole fields with $l=3$ and $l=2$.

\begin{figure}[tbp]
\centering
\includegraphics[width=3.4in]{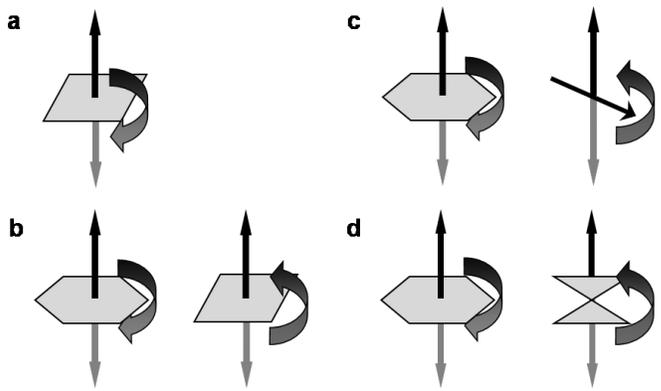}
\caption{(Color online). Different vortex pump protocols compared:
from the top to the bottom. The parallelogram and hexagon denote,
respectively, 2D quadrupole and hexapole B-fields, and the stacked
pyramids denote a 3D quadrupole B-field. The initial $z$-bias is
denoted by the solid black line arrow that is flipped to the
opposite direction denoted by gray line arrow in the first step, the
slanted solid black line arrow in the right figure of (c) refers to
the transverse bias. (a) The original theoretical protocol of Ref.
\cite{Machida}, which has been experimentally demonstrated
\cite{emit1,emit2,japan} can pump only once. (b) The continuous pump
protocol of Ref. \cite{pump}. It works, but the condensate vorticity
does not monotonically change; (c) and (d) are the optimal protocols
we discuss. They are capable of continuous and monotonic changes to
the vorticity of a condensate. They are the optimal protocols based
on the manipulation of B-fields.} \label{fig5}
\end{figure}

\begin{figure}[tbp]
\centering
\includegraphics[width=3.4in]{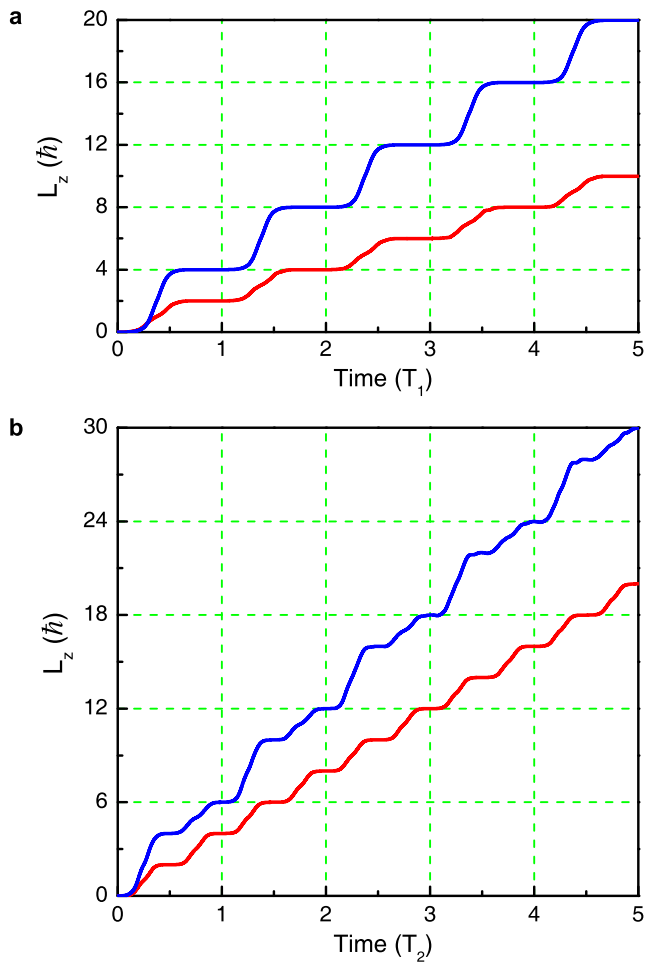}
\caption{(Color online). The continuously increasing vorticity of
a condensate for the optimal protocols. (a) with only
$\vec{B}_{l}^{(I)}(\vec{r})$ for a hexapole ($l=3$ the upper blue
curve) and a quadrupole ($l=2$ the lower red curve). $T_1=40$ ms
is the time of flipping. In the first 30 ms
$\vec{B}_{l}^{(I)}(\vec{r})$ is turned on to gain vorticity, while
no vorticity is gained in the following 10 ms when a transverse
bias along the $x$-axis is turned on to reset the internal states
to be as they were initially. (b) with both B-fields
$\vec{B}_{l}^{(I)}(\vec{r})$ ($l=3$ for the upper blue curve and
$l=2$ for the lower red curve) and $\vec{B}^{(II)}(\vec{r})$
available and used for the first and second steps, respectively, of
each cycle. $T_2=60$ ms is the total time for each cycle. In the
first 30 ms $\vec{B}_{l}^{(I)}(\vec{r})$ is present and in the
next 30 ms $\vec{B}^{(II)}(\vec{r})$ is turned on while
$\vec{B}_{l}^{(I)}(\vec{r})$ is turned off. The vorticity
increases at all times in this case. The dashed lines in two
figures are guides for the eye.} \label{fig6}
\end{figure}

Finally, in light of our analysis above, the protocol
devised in Ref. \cite{pump}, which uses only the field of
$\vec{B}_{l}^{(I)}(\vec{r})$ with $l=3$ and $l=2$, respectively, in
the first and the second steps of each cycle, is clearly not
optimal. In the first step, the condensate gains a vorticity of
$4\hbar$, while in the second step, it gains a vorticity of
$-2\hbar$, or loses $2\hbar$. In the end, the net gain is only a
$2\hbar$ per cycle.
We summarize the various vortex pump protocols based on the
manipulations with external B-fields in Fig. \ref{fig5}.

In actual experiments, many complications could arise that make the
selection of the optimal vortex pump protocol an entirely
system-dependent matter. For instance, the loss of atoms and
quantum coherence due to dissipation or
decoherence strongly limit the lifetimes of all currently available
atomic superfluids. If the criterion for the optimal protocol is
defined as the maximal amount of vorticity gained during a fixed
time, our analysis above then points to a clear winner: the repeated
applications with only the hexapole field
$\vec{B}_{3}^{(I)}(\vec{r})$ as evidenced by the results in the
illustrated figures above.

Before concluding, we want to point out that $D_z$ remains conserved
within our formulation using the exact eigen-energies for the three
Zeeman states, which is indeed confirmed in numerical simulations.
This is easily understood because the effective Zeeman interaction
$H^{ZM}=E_0-\eta_0\vec{F}\cdot\vec{B}/B+\delta(\vec{F}\cdot\vec{B})^2/B^2$
with $\eta_0=(E_{-1}-E_1)/2$ and $\delta=(E_{1}+E_{-1}-2E_0)/2$
commutes with $D_z$. In addition, $L_z$ is also conserved during
the $10$ ms intervals when only a uniform B-field is present.

\section{Conclusion}

In conclusion, we have proposed and demonstrated numerically a simple yet
efficient vortex pump protocol capable of continuously increasing or
decreasing the vorticity of a condensate through repeated
manipulations of external B-fields: a bias along the axial
direction, a 2D quadrupole field or hexapole field, and a 3D
quadrupole field or a transverse bias field. Based on a general
consideration of the geometrical properties of static B-fields, we
have shown that the above choices of B-fields are the only
possibilities capable of continuously controlling angular momentum
vorticity in a spinor condensate through the manipulation of
B-fields. In Fig. \ref{fig6} we compare the two optimal protocols
for continuously increasing vorticity for the first five
repetitions. The vorticity is seen increasing to $20 \hbar$ and $10
\hbar$ by using only $\vec{B}_{l}^{(I)}(\vec{r})$ for a hexapole
($l=3$) and a quadrupole ($l=2$), respectively, in (a), and to $30
\hbar$ and $20 \hbar$ by combining both $\vec{B}_{l}^{(I)}(\vec{r})$
and $\vec{B}^{(II)}(\vec{r})$ for $l=3$ and $l=2$, respectively, in
(b). To assure adiabaticity throughout the B-field manipulations,
each cycle has to be engineered differently in the two protocols.
For the first protocol as shown in (a), the duration for each cycle,
or its period is $T_1=40$ ms. During the first 30 ms we use
$\vec{B}_{l}^{(I)}(\vec{r})$ to increase vorticity, and in the
following 10 ms the transverse bias along the $x$-axis is turned on
to reset to the initial internal state through a second $z$-bias
flip. For the second protocol of (b), the period required becomes
$T_2=60$ ms. During the first 30 ms $\vec{B}_{l}^{(I)}(\vec{r})$ is
present and in the following 30 ms $\vec{B}^{(II)}(\vec{r})$ is
turned on while $\vec{B}_{l}^{(I)}(\vec{r})$ is turned off.

Our discussions indicate that for $l=2$, the
second protocol is more efficient if a maximal amount of vorticity
is sought. While for $l=3$, our results show that the
efficiencies for the two protocols are about the same due to the
different time requirements for adiabaticity. Based on the
experience from the numerical simulations and from experimental
works, it seems more practical to use the first protocol. The
reason is rather simple, it is always more difficult to
have two types of magnetic traps (from two sets of current
carrying coils) working together. The first protocol requires only
one type of multipole 2D B-field $\vec{B}_{l}^{(II)}(\vec{r})$,
that can be realized on an atom chip \cite{fortagh}, using a
three- or four-wire structure for a 2D QT \cite{Lesanovsky}, or a
five-wire structure for a 2D hexapole trap \cite{Esteve}.

We demonstrate above the practical limit
of reaching $30\hbar$ per atom using our protocol,
which is already much larger
than anything reported experimentally through the stirring
of a condensate. We do not know what the maximum limit is.
In fact, the exact value of this maximum is perhaps not
so important either for two reasons:
1), our theory is mean field and may well break down
at such larger angular momentum per atom because
quantum correlations become essential;
2), we are also limited by the computational resources
available for such large scale simulations.

The prospect for realizing the proposed vortex pump mechanism
provides new impetus to active pursuits of quantum simulations of
strongly interacting many body electronic systems
in terms of cold atoms \cite{papen,kav1,kav2,
zol,hofr,wilkin,mac,baym,baranov,jila0,ens0,mit0,ryu,morris}.
It opens new avenues of theoretical and experimental research
into the coupling between the internal and motional degrees of
freedom and shines new light on relevant topics in
spintronics studies \cite{kato}.

Finally we note that in all our protocols discussed above,
there always exist a conserved quantity $D_z$ for the ideal case
of the model system. To create a vortex, one simply
needs to adiabatically flip the B-field to transfer all atoms
from the weak field seeking Zeeman state $|-1\rangle_z$ to
$|+1\rangle_z$, or the reverse. An obvious connection can be
made with the famous Einstein-de Haas effect to deepen
the understanding of our protocol and to possibly extend
to the systems of dipolar quantum gases \cite{ueda,pol}.
During our extensive numerical
simulations, we find that a small B-field gradient $B'$ or $B'_h$,
combined with a stronger optical plug enforces the stability of the
high vorticity states as generated. As is well known
high-order multipole B-field traps are usually not ideal
candidates for magnetic traps because they are too weak to trap atoms.
As far as vortex pump is concerned, however, a
hexapole trap is preferred over a quadrupole trap because
we are making use the topological structure of the B-field,
rather than its spatial dependence when trapping is considered.

To help with further experimental effort, we suggest a feasible
implementation scheme on an atom chip. As is well known, an IPT can
be realized by combing a chip based $Z$-type trap with a constant
B-field \cite{fortagh}. This type of IPT is not an ideal setup for
implementing our protocols because the locations of trap zeros at
the vanishing B-field change following increasing or decreasing
currents in the on chip wires. The IPTs from the more stable three-
or four-wire structures, on the other hand, are more promising
candidates as evidenced by their demonstrated successes and
stabilities in our chip matter wave manipulations \cite{Lesanovsky}.
For the three-wire structure, three equally spaced parallel wires
lay on the surface of an atom chip, with the current in the middle
wire opposite to the other two side wires. The four-wire structures
are analogously configured. Following similar ideas, a
two-dimensional hexapole trap can be constructed by five wires
\cite{Esteve}. Additionally these chip-based traps can be combined
with optical traps, such as an optical plug, to realize a rich
variety of confinement geometries. The force of gravity can be made
to align along any of the symmetry axes of the two-dimensional
multipole trap or compensated with an additional optical
confinement.

This work is supported by US NSF, and NSFC and MOST of China.
C.R. acknowledges support from ARO.

\end{document}